\documentclass[%
reprint,
 amsmath,amssymb,
 aps,
]{revtex4-1}

\usepackage{graphicx}
\usepackage{dcolumn}
\usepackage{bm}
\usepackage[section]{placeins}
\usepackage{color}
\usepackage{mathrsfs}
\usepackage{ulem}
\usepackage{hyperref}
\usepackage{braket}
\usepackage{siunitx}

\begin{document}

\title{Probing the Meissner effect in single crystals of $\mathbf{Bi_2Sr_2Ca_2Cu_3O_{10+\delta}}$ via wide-field quantum microscopy under high pressure}

\author{Masahiro Ohkuma$^{1}$}
\email{okuma.m.36c3@m.isct.ac.jp}
\author{Ryo Matsumoto$^{2}$}
\author{Shintaro Adachi$^{3}$}
\author{Shinobu Onoda$^{4}$}
\author{Takao Watanabe$^{5}$}
\author{Kenji Ohta$^{6}$}
\author{Yoshihiko Takano$^{2}$}
\author{Keigo Arai$^{1}$}

\affiliation{$^{1}$School of Engineering, Institute of Science Tokyo, Yokohama 226-8501, Kanagawa, Japan}
\affiliation{$^{2}$Research Center for Materials Nanoarchitectonics (MANA), National Institute for Materials Science, Tsukuba 305-0047, Ibaraki, Japan}
\affiliation{$^{3}$Department of Mechanical and Electrical Systems Engineering, Kyoto University of Advanced Science (KUAS), Kyoto 615-8577, Kyoto, Japan}
\affiliation{$^{4}$National Institutes for Quantum Science and Technology (QST), Takasaki 370-1292, Gunma, Japan}
\affiliation{$^{5}$Graduate School of Science and Technology, Hirosaki University, Hirosaki 036-8561, Aomori, Japan}
\affiliation{$^{6}$Department of Earth and Planetary Sciences, Institute of Science Tokyo, Meguro 152-8551, Tokyo, Japan}

\date{\today}

\begin{abstract}
We investigated the pressure dependence of the superconducting transition temperature ($T_{\rm c}$) in optimally doped Bi$_2$Sr$_2$Ca$_2$Cu$_3$O$_{10+\delta}$ (Bi-2223) single crystals using different pressure-transmitting media.
Previous high-pressure studies have reported conflicting behaviors, ranging from a resurgence of $T_{\rm c}$ of optimally doped Bi-2223 in fluid media to an insulating-like transition in solid media.
However, a direct comparison of the effects of different pressure-transmitting media is lacking.
Here, we employed wide-field quantum microscopy based on nitrogen-vacancy centers to probe the magnetic response under high pressure, utilizing cBN and KBr as media.
We observed that a diamagnetic response near 70 K, indicative of the superconducting transition, persisted up to 23 GPa in KBr, whereas it disappeared above 11 GPa and 70 K in cBN.
These results demonstrate the high sensitivity of Bi-2223 to the pressure environment and highlight the critical role of hydrostatic pressure in cuprate superconductors.
\end{abstract}

\maketitle

High-pressure techniques are powerful tools for directly modulating material structures, enabling the exploration of emergent physical properties and the discovery of novel materials \cite{shimizuPressureinducedSuperconductivityElemental2005, yamanakaSiliconClathratesCarbon2010, maoSolidsLiquidsGases2018}.
In characterizing physical properties under these extreme high-pressure conditions, on the order of several tens of GPa or above, achieving isotropic compression is as crucial as the pressure magnitude.
For instance, in superconducting phases that were recently discovered in nickel oxides under high pressure, maintaining isotropic pressure is known to be advantageous for stabilizing superconducting phases \cite{sunSignaturesSuperconductivity802023a, sakakibaraTheoreticalAnalysisPossibility2024, wangPressureInducedSuperconductivityPolycrystalline2024, zhangHightemperatureSuperconductivityZero2024b, zhuSuperconductivityPressurizedTrilayer2024}.
In the case of Bi-based cuprates, the pressure dependence of the superconducting transition temperature ($T_{\rm c}$) initially exhibits a dome-shaped behavior \cite{chenHighpressurePhaseDiagram2004a, chenEnhancementSuperconductivityPressuredriven2010, adachiPressureEffectBi22122019, dengHigherSuperconductingTransition2019a, matsumotoCrystalGrowthHighPressure2021, zhouQuantumPhaseTransition2022, markProgressProspectsCuprate2022}.
In $\rm {Bi_2Sr_2Ca_2Cu_3O_{10+\delta}}$ (Bi-2223), however, while $T_c$ exhibits an upturn at higher pressures when a fluid pressure-transmitting medium is used, the use of a solid medium leads to an insulating-like transition and the subsequent disappearance of the zero-resistance state \cite{chenEnhancementSuperconductivityPressuredriven2010, adachiPressureEffectBi22122019}.
Although high-pressure structural analyses suggest that these discrepancies stem from the pressure-transmitting media dependence of structural modifications \cite{markStructureEquationState2023}, there have been no direct comparative studies evaluating the superconducting properties using different pressure-transmitting media in the high-pressure region.

\begin{figure}[htb!]
\centering\includegraphics[clip,width=1\columnwidth]{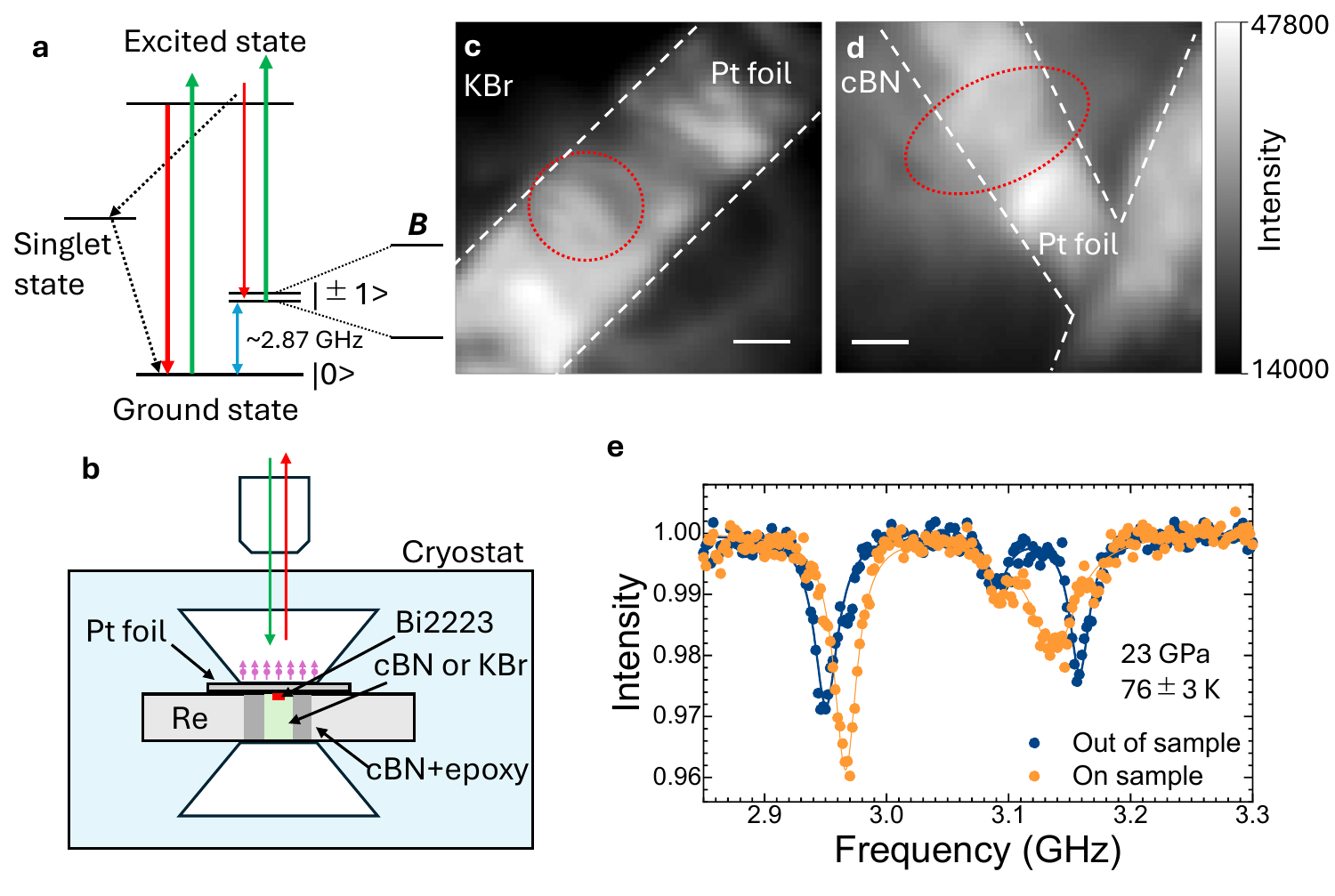}
\caption{\label{f1} (a) Energy diagram of the NV center.
(b) Overview of the experimental setup.
(c) and (d) Wide-field photoluminescence image of the NV centers inside DAC using (c) KBr and (d) cBN. The red circles indicate the position of Bi-2223. Each scale bar represent 20 $\mu$m.
(e) ODMR spectra of NV centers above and outside the superconducting region.}
\end{figure}

$T_{\rm c}$ under high pressure is mainly evaluated using electrical transport and magnetic measurements.
In transport measurements, the need for electrical contacts and wiring makes experiments using fluid-pressure-transmitting media difficult.
In conventional magnetic measurements, the signal amplitude scales directly with the sample volume, which often limits the pressure range \cite{mitoMagnetizationMeasurementsUsing2025}.
Consequently, existing approaches have difficulty performing systematic studies across pressure-transmitting media ranging from solids to fluids and even gases.
Recently, the negatively charged nitrogen-vacancy (NV) center in diamond has emerged as a highly sensitive nanoscale probe for sensing, including under extreme conditions \cite{balasubramanianNanoscaleImagingMagnetometry2008a, degenScanningMagneticField2008, mazeNanoscaleMagneticSensing2008, taylorHighsensitivityDiamondMagnetometer2008a, toyliMeasurementControlSingle2012a, dohertyElectronicPropertiesMetrology2014, schaefer-nolteDiamondbasedScanningProbe2014, liuCoherentQuantumControl2019, hsiehImagingStressMagnetism2019, lesikMagneticMeasurementsMicrometersized2019, yipMeasuringMagneticField2019, scheideggerScanningNitrogenvacancyMagnetometry2022a, bhattacharyyaImagingMeissnerEffect2024, wangImagingMagneticTransition2024}.
The NV center has an electronic spin of $S=1$, whose spin state can be initialized and read out optically and coherently manipulated using resonant microwave fields (Fig.~\ref{f1}a) \cite{oortOpticallyDetectedSpin1988, gruberScanningConfocalOptical1997a}.
This technique is known as optically detected magnetic resonance (ODMR).
By integrating a thin layer of NV centers into a diamond anvil cell (DAC), it is possible to perform high-spatial-resolution imaging of physical quantities, such as magnetic fields and stress, directly at the sample interface \cite{hsiehImagingStressMagnetism2019, lesikMagneticMeasurementsMicrometersized2019}.
In contrast to conventional magnetometry, which typically involves signals from the entire pressure cell, this quantum sensing platform enables the direct observation of local magnetic fields.
This spatially resolved approach offers a distinct advantage for the direct detection of the Meissner effect in micrometer-sized samples, even above 100 GPa \cite{bhattacharyyaImagingMeissnerEffect2024, chenImagingMagneticFlux2025a, haoDiamondQuantumSensing2025}.

In this study, we conducted wide-field quantum microscopy to investigate the superconducting properties of Bi-2223 single crystals under high pressure using different pressure-transmitting media KBr and cBN.
Both KBr and cBN are solids; however, hydrostaticity can differ due to different of the mechanical properties.
We spatially resolved the diamagnetic signal via the NV center and observed a pronounced dependence on the pressure-transmitting medium: a clear diamagnetic response persisted up to 23 GPa in KBr near 70 K, whereas it was suppressed above 11 GPa in cBN.
These results highlight the critical role of hydrostatic pressure in cuprate superconductors.

Pressure was applied using a DAC made of CuBe.
An overview of the experimental setup is shown in Fig.~\ref{f1}b.
We used a type Ib diamond anvil with a culet diameter of 300~$\mu$m and a $\{111\}$-oriented culet surface (SYNTEK Co., Ltd.).
To create NV centers, ${}^{12}\mathrm{C}^{+}$ ions were implanted at an energy of 30~keV with a dose of $5\times10^{12}~\mathrm{cm}^{-2}$, and the diamond anvil was subsequently annealed in vacuum at $\SI{1000}{\degreeCelsius}$ for 2~h.
A rhenium gasket was pre-indented to a thickness of 50~$\mu$m, and a hole with a diameter of 250~$\mu$m was drilled.
To prevent electrical contact between the microwave waveguide and the rhenium gasket, the hole was filled with a cBN--epoxy mixture, pre-indented, and then the sample chamber with a diameter of 100~$\mu$m was drilled.
A Pt foil with thickness of 5~$\mu$m and Bi-2223 were placed on the diamond anvil.
The sample chamber was filled with cBN or KBr powder as the pressure-transmitting medium.
Wide-field photoluminescence images are shown in Figs.~\ref{f1}c and d.
Pressure was estimated from the diamond Raman shift measured at room temperature using a spectrometer (inVia, Renishaw) \cite{akahamaHighpressureRamanSpectroscopy2004}.
The results of Raman spectroscopy measurements are shown in supplemental material.
The Bi-2223 single crystal was grown by the traveling-solvent floating-zone method, and as-grown Bi-2223 single crystals were annealed at $\SI{500}{\degreeCelsius}$ under a flowing 100$\%$ oxygen gas (oxygen partial pressure of 1 atm) to obtain optimally doped samples with $T_{\rm c}\sim110$ K \cite{fujiiSinglecrystalGrowthBi2Sr2Ca2Cu3O10+d2001, fujiiDopingDependenceAnisotropic2002, adachiSinglecrystalGrowthUnderdoped2015, adachiUnscalingSuperconductingParameters2015}.

Low-temperature wide-field ODMR measurements were performed using a home-built optical setup integrated with a custom-built Gifford--McMahon refrigerator.
The temperature was measured using a Cernox thermometer mounted on a stage in thermal contact with the DAC and monitored with a temperature controller (Lake Shore, Model 335).
A green excitation laser (MLL-S-532B, CNI laser) was focused onto the sample through an objective lens (M-PLAN APO 7.5X, Mitsutoyo), and the red fluorescence was collected by the same objective.
The photoluminescence was filtered by a long-pass filter and imaged onto an EMCCD camera (Andor, iXon Ultra 897).
Microwaves were generated using a signal generator (SynthHD, Windfreak) and amplified using a power amplifier (ZHL-16W-43$+$, Mini-circuits).
To minimize mechanical vibrations, the cold head was deactivated during data acquisition so that the temperature increased during the measurement.
The temperature was defined as the mean of the temperatures measured immediately before and after each acquisition, while the temperature uncertainty was taken as the full range between these two values.
Each ODMR spectrum was acquired within up to 15 min.
A magnetic field was applied along the $\langle111\rangle$ direction using a coil wound around the DAC.

\begin{figure}[htb!]
\centering\includegraphics[clip,width=1\columnwidth]{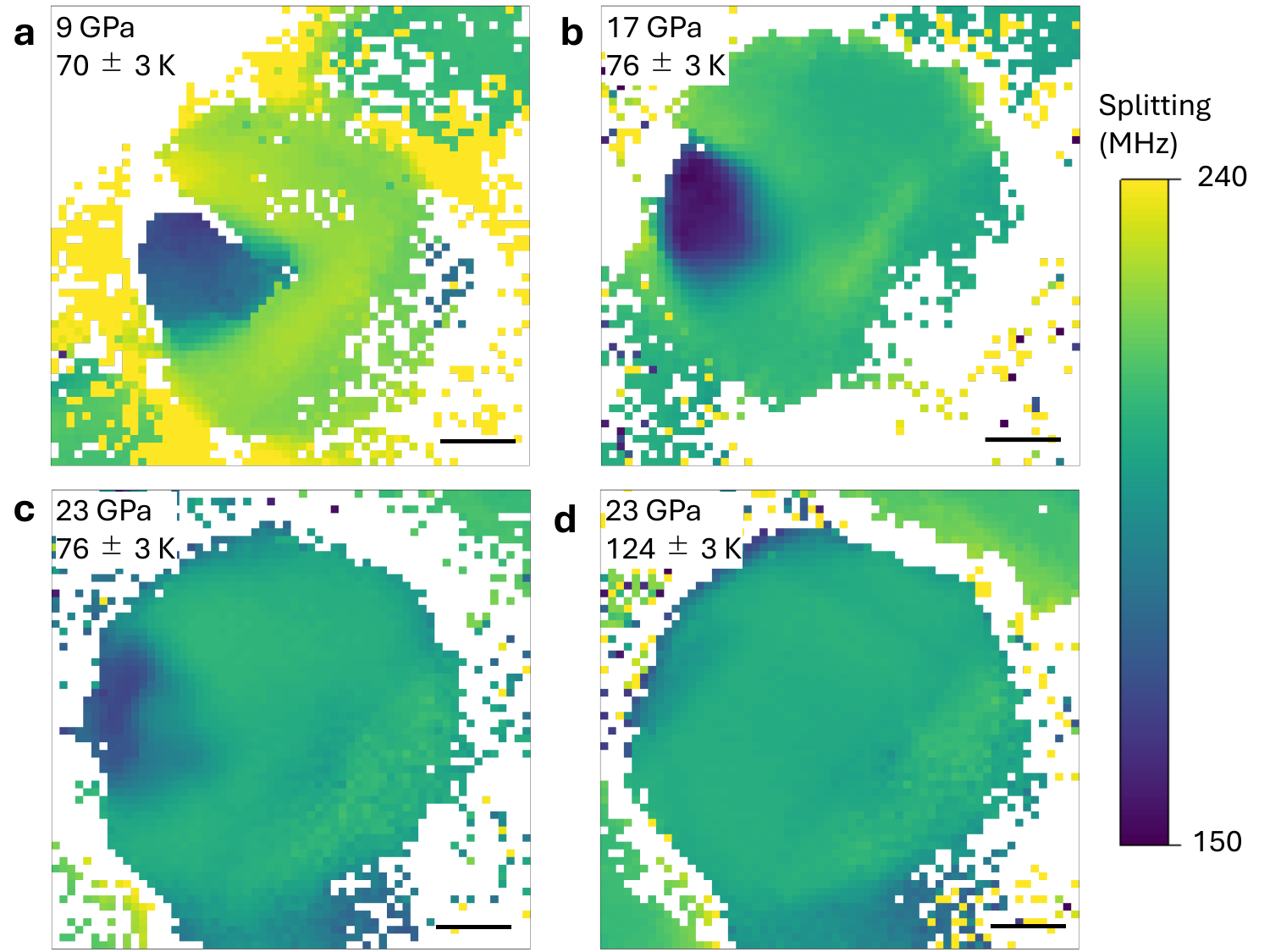}
\caption{\label{f2} Map of the ODMR splitting using KBr as the pressure-transmitting medium. (a)--(c) Results below $T_{\rm c}$ with different pressures. (d) Result above $T_{\rm c}$ at 23 GPa. Each scale bar represents 20 $\mu$m.}
\end{figure}

The spin sublevels of the NV center undergo Zeeman splitting when a magnetic field is applied.
When a superconductor is placed on top of the NV layer, the local magnetic field sensed by the NV centers is reduced below $T_\mathrm{c}$ owing to magnetic flux expulsion (the Meissner effect).
Consequently, the ODMR splitting measured above the superconducting region becomes smaller than that measured outside the superconducting region.
Figure~\ref{f1}e shows representative ODMR spectra obtained below $T_\mathrm{c}$.
After cooling the sample to below 20~K, we performed zero-field cooling (ZFC) and acquired ODMR spectra during the subsequent warming process.
A magnetic field of approximately 4~mT was applied.
The pressure was 23~GPa, the temperature was $76\pm3$~K, and KBr was used as the pressure-transmitting medium (the corresponding wide-field image is shown in Fig.~\ref{f2}c).
The spectrum exhibits three resolved resonance peaks: the two outer resonances are attributed to NV centers, whose axes are nearly aligned with the applied field, whereas the central feature originates from NV centers with other crystallographic orientations.
The fourth resonance expected from these non-aligned orientations was not separately resolved because it overlapped with the high-frequency $\langle111\rangle$ resonance within the linewidth.
Below $T_\mathrm{c}$, the splitting associated with the field-aligned NV centers is reduced above the superconducting region, indicating that Meissner screening persists even at 23~GPa. 
In the following section, we present the wide-field ODMR results.
In each measurement, the pressure was increased at room temperature, and ODMR spectra were acquired during the warming process after ZFC.

Figure~\ref{f2} shows the results obtained using KBr as the pressure-transmitting medium.
Data points for which the half-width of the 95$\%$ confidence interval of the fitted resonance frequencies of the two outer peaks exceeded 5 MHz were excluded from the analysis.
ODMR signals were clearly observable only in the vicinity of the Pt foil.
In regions where the ODMR features became barely discernible, the fitted peak positions exhibited large uncertainties and were therefore excluded.
We attribute the reduced ODMR visibility mainly to weaker microwave driving and an insufficient signal-to-noise ratio owing to limited averaging.
In Figs.~\ref{f2}a--c, a clear diamagnetic signal was observed below 80~K.
In contrast, the diamagnetic signal disappeared well above $T_\mathrm{c}$, as shown in Fig.~\ref{f2}d.
A comparison of Figs.~\ref{f2}b and c reveals that the diamagnetic signal is smaller in Fig.~\ref{f2}c, suggesting that $T_\mathrm{c}$ decreases with increasing pressure.

\begin{figure}[htb!]
\centering\includegraphics[clip,width=1\columnwidth]{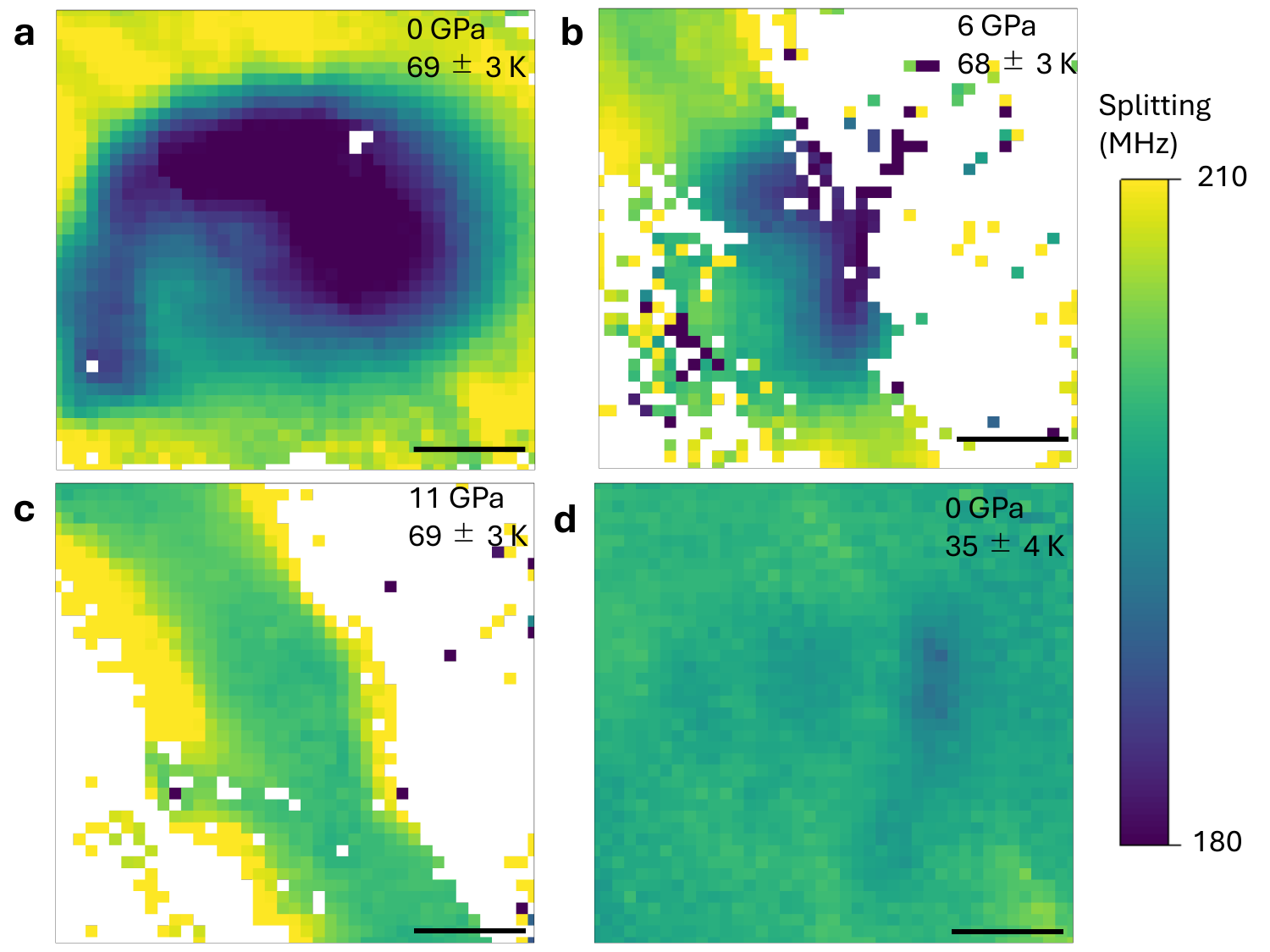}
\caption{\label{f3} Map of the ODMR splitting using cBN as the pressure-transmitting medium. (a) and (b) Results below $T_{\rm c}$ with different pressures. (c) Result above $T_{\rm c}$ at 11 GPa. (d) Result after pressure releasing from 19 GPa. Each scale bar represents 20 $\mu$m.}
\end{figure}

Next, we performed experiments using cBN as the pressure-transmitting medium.
Figures~\ref{f3}a--c show images acquired at temperatures of approximately 70~K.
At 0~GPa, a diamagnetic signal was observed over the entire sample area.
However, analyzable ODMR spectra were obtained only in the vicinity of the Pt foil upon pressurization.
A diamagnetic signal was detected directly beneath the Pt foil at 6~GPa, whereas no clear diamagnetic signal was observed at 11~GPa above 70 K.
We further increased the pressure up to 19~GPa; however no clear diamagnetic signal was detected above 30~K with our current setup.
After pressure release, a very weak diamagnetic signal reappeared, as shown in Fig.~\ref{f3}d.
A reduction in the diamagnetic signal after pressurization has also been observed in magnetic measurements of cuprate superconductors \cite{mitoUniaxialStrainEffects2016, mitoUniaxialStrainEffects2017}.

\begin{figure}[htb!]
\centering\includegraphics[clip,width=1\columnwidth]{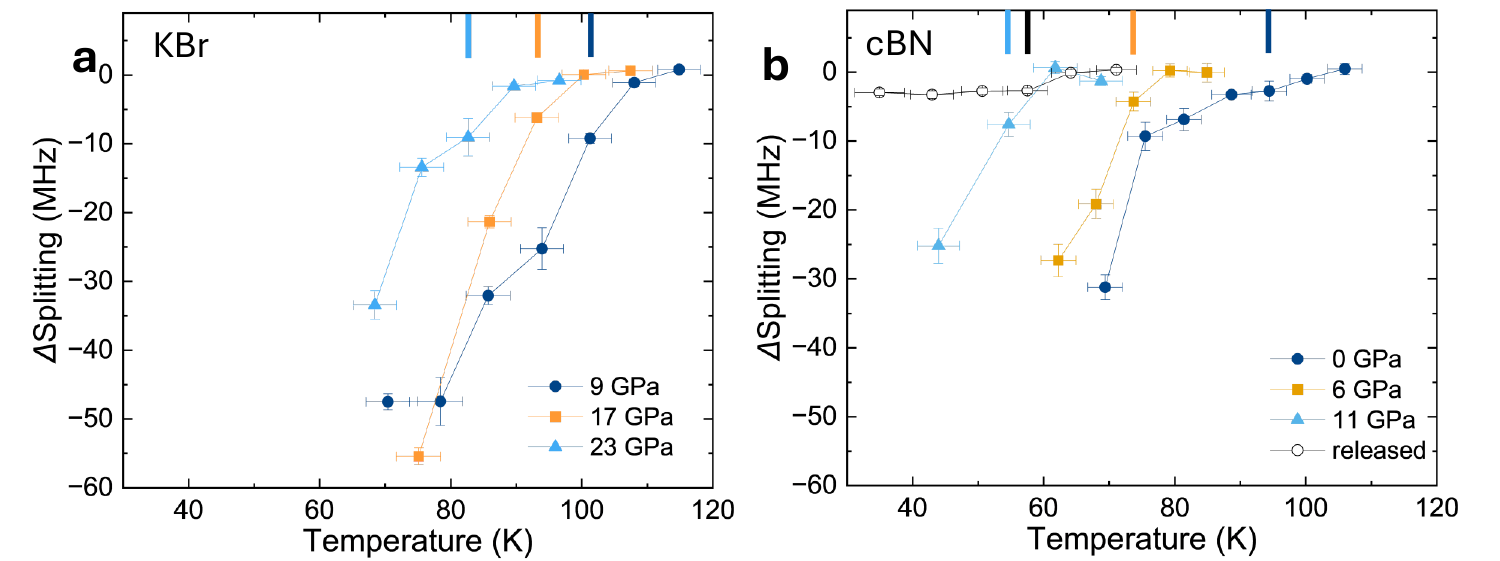}
\caption{\label{f4} Temperature dependence of ODMR splitting. Results using (a) KBr and (b) cBN as the pressure-transmitting media. The shift was calculated as the difference between the mean values of pixels on the sample and those in the background, with error bars representing the combined uncertainty derived from the standard deviations of both pixel groups. Bars represents the onset temperature where the diamagnetic signals are observed.}
\end{figure}

Figure~\ref{f4} summarizes the temperature dependence of the ODMR splitting.
The shift was calculated as the difference between the mean values of the pixels on the sample and those in the background, with error bars representing the combined uncertainty derived from the standard deviations of both pixel groups.
In the measurements using KBr, the onset of the diamagnetic signal shifts toward lower temperatures with increasing pressure, suggesting a decrease in $T_{\mathrm{c}}$.
Conversely, in cBN, the non-hydrostatic pressure significantly suppresses superconductivity.
At 11~GPa, the diamagnetic response was clearly visible; however, at 19~GPa, we could not resolve clear a diamagnetic signal down to 30~K in our setup.
This behavior is consistent with previous electrical measurements using cBN, which reported a much lower zero-resistance temperature and insulating-like resistance behavior.
Notably, after releasing the pressure from 19~GPa, a diamagnetic signal re-emerged below 60~K.
However, the diamagnetic signal was observed at a lower temperature than in the initial state, and its magnitude did not fully recover.
This irreversibility suggests that the $T_{\rm{c}}$ suppression in cBN is not solely driven by lattice compression; rather, extrinsic effects, such as lattice distortion or grain refinement, likely play a significant role.

Our results indicate that when cBN is used as a pressure-transmitting medium, superconductivity can be strongly suppressed and may evolve toward an insulating-like behavior.
Similar tendencies have also been reported for Bi$_2$Sr$_2$CaCu$_2$O$_{8+\delta}$ (Bi-2212), suggesting that the behavior of Bi-based cuprates under high pressure may depend sensitively on the pressure environment.
For example, an insulating-like behavior has been reported when using cBN, whereas with silicone oil, $T_{\rm{c}}$ is reported to increase again above approximately 40 GPa \cite{dengHigherSuperconductingTransition2019a, matsumotoCrystalGrowthHighPressure2021}.
Related behavior has also been discussed for experiments using NaCl as a pressure-transmitting medium, where an insulating-like change appears upon pressurization \cite{zhouQuantumPhaseTransition2022}.
Notably, in the report, no obvious cracks were observed after pressurization, and zero resistance as well as metallic conductivity were recovered after pressure release, implying that the insulating-like behavior is not caused by extrinsic effects \cite{zhouQuantumPhaseTransition2022}.
While structural analyses under high pressure emphasize the importance of hydrostaticity in cuprate superconductors, a unified understanding of why different pressure media can lead to qualitatively different outcomes remains elusive \cite{markStructureEquationState2023}.

To resolve these discrepancies, a local and non-contact probe that can operate under high-pressure environments is highly desirable.
Wide-field quantum microscopy under high pressure is well suited to address this issue because it enables contact-free, spatially resolved measurements of magnetic responses in micrometer-scale samples.
This capability opens a route to systematically vary the pressure-transmitting medium from solids to gases and track how superconducting signatures evolve under controlled pressure environments.
In particular, the spatial information provided by NV-based magnetometry may help disentangle contributions from hydrostaticity versus non-hydrostatic stress and local strain, which are difficult to separate in conventional bulk magnetometry or transport measurements.
Furthermore, if advanced quantum-metrology protocols, such as quantum noise spectroscopy, can be extended to high-pressure conditions, it is expected to provide deeper insights into the pressure-driven transition \cite{schafer-nolteTrackingTemperatureDependentRelaxation2014, liuQuantumNoiseSpectroscopy2025}.

Given that the superconductivity of cuprates can be sensitive to the pressure environment, it is crucial to distinguish intrinsic pressure effects from those arising from non-hydrostatic pressure.
Previous studies on Hg-based cuprates under hydrostatic pressure using a cubic anvil cell (CAC) have successfully demonstrated zero resistance up to approximately 22 GPa \cite{takeshitaZeroResistivity1502013, yamamotoHighPressureEffects2015}.
Furthermore, comparative magnetic measurements using SQUID have revealed that $T_{\mathrm{c}}$ is maximally enhanced under hydrostatic pressure compared to uniaxial strain conditions \cite{mitoUniaxialStrainEffects2017}, underscoring the critical role of hydrostatic pressure in cuprate superconductors.
However, achieving higher pressures in CAC presents significant technical challenges.
Although DAC can generate higher pressures, SQUID magnetometry becomes difficult under high pressures where the sample space is restricted owing to insufficient magnetic signals.
In this context, NV-based magnetometry offers a distinct advantage: it provides high sensitivity even for micrometer-sized samples and does not require electrical contacts.
This enables the use of fluid-pressure-transmitting media within the DAC, ensuring a hydrostatic environment even at ultra-high pressures.
Consequently, this technique paves the way for extending the pressure-$T_{\mathrm{c}}$ phase diagram and exploring the potential secondary superconducting dome in the unexplored high-pressure regime \cite{nokelainenSecondDomeSuperconductivity2024}.

More broadly, wide-field quantum microscopy under high pressure has emerged as a contact-free, spatially resolved tool for probing quantum phenomena in micrometer-scale samples, where conventional magnetometry and transport measurements often face fundamental limitations.
Because many quantum materials, such as van der Waals layered compounds and nickelate superconductors, are fragile and/or highly sensitive to non-hydrostatic stress, the ability to operate in a DAC without electrical contacts and with improved hydrostatic pressure environments is particularly valuable.
This approach provides a practical route to extend pressure–temperature phase diagrams to higher pressures under more hydrostatic conditions and to explore emergent quantum phenomena.
In the future, combining NV-based magnetometry with fluid pressure-transmitting media and complementary in-situ diagnostics (e.g., stress mapping and low-temperature pressure calibration) will further strengthen the quantitative interpretation and enable systematic studies across a broad range of quantum materials.

In conclusion, we demonstrated wide-field quantum microscopy using the NV center of a Bi-2223 single crystal under high pressure.
By directly imaging the local diamagnetic response associated with the Meissner effect, we observed a pronounced dependence on the pressure-transmitting medium: a clear superconductivity-related diamagnetic signal persisted up to 23 GPa in KBr near 70 K, whereas it was suppressed above 11 GPa and 70 K in cBN.
These results underscore the critical role of hydrostatic pressure in establishing reliable pressure-$T_{\mathrm{c}}$ relationships in Bi-based cuprate superconductors.

\section*{DATA AVAILABILITY}
The data that support the findings of this study are available from the corresponding author upon reasonable request.

\begin{acknowledgments}
 This work was supported by JSPS KAKENHI Grant numbers JP23K26528, JP23KK0267, JP24KJ1035, JP25K01508. This work was also supported by JST ASPIRE Grant number JPMJAP24C1. M.O. receives funding from JSPS Grant-in-Aid for JSPS Fellows Grant number JP24KJ1035.
\end{acknowledgments}

\bibliography{NV}

@misc{liuQuantumNoiseSpectroscopy2025,
  title = {Quantum Noise Spectroscopy of Superconducting Dynamics in Thin-Film {Bi$_2$Sr$_2$CaCu$_2$O$_{8+\delta}$}},
  author = {Liu, Zhongyuan and Gong, Ruotian and Kim, Jaewon and Diessel, Oriana K. and Xu, Qiaozhi and Rehfuss, Zackary and Du, Xinyi and He, Guanghui and Singh, Abhishek and Eo, Yun Suk and Henriksen, Erik A. and Gu, G. D. and Yao, Norman Y. and Machado, Francisco and Ran, Sheng and Chatterjee, Shubhayu and Zu, Chong},
  year = 2025,
  month = feb,
  number = {arXiv:2502.04439},
  eprint = {2502.04439},
  primaryclass = {cond-mat},
  publisher = {arXiv},
  doi = {10.48550/arXiv.2502.04439},
  archiveprefix = {arXiv}
}

@article{mitoMagnetizationMeasurementsUsing2025,
  title = {Magnetization Measurements Using {{SQUID}} with Diamond Anvil Cells under Extremely High Pressure},
  author = {Mito, Masaki and Hamada, Masayoshi},
  year = 2025,
  month = jul,
  journal = {Appl. Phys. Rev.},
  volume = {12},
  number = {3},
  pages = {031310},
  issn = {1931-9401},
  doi = {10.1063/5.0229926}
}

@article{schafer-nolteTrackingTemperatureDependentRelaxation2014,
  title = {Tracking {{Temperature-Dependent Relaxation Times}} of {{Ferritin Nanomagnets}} with a {{Wideband Quantum Spectrometer}}},
  author = {{Sch{\"a}fer-Nolte}, Eike and Schlipf, Lukas and Ternes, Markus and Reinhard, Friedemann and Kern, Klaus and Wrachtrup, J{\"o}rg},
  year = 2014,
  month = nov,
  journal = {Phys. Rev. Lett.},
  volume = {113},
  number = {21},
  pages = {217204},
  publisher = {American Physical Society},
  doi = {10.1103/PhysRevLett.113.217204}
}

@article{adachiSinglecrystalGrowthUnderdoped2015,
  title = {Single-Crystal {{Growth}} of {{Underdoped Bi-2223}}},
  author = {Adachi, S. and Usui, T. and Takahashi, K. and Kosugi, K. and Watanabe, T. and Nishizaki, T. and Adachi, T. and Kimura, S. and Sato, K. and Suzuki, K. M. and Fujita, M. and Yamada, K. and Fujii, T.},
  year = 2015,
  month = jan,
  journal = {Physics Procedia},
  series = {Proceedings of the 27th {{International Symposium}} on {{Superconductivity}} ({{ISS}} 2014) {{November}} 25-27, 2014, {{Tokyo}}, {{Japan}}},
  volume = {65},
  pages = {53--56},
  issn = {1875-3892},
  doi = {10.1016/j.phpro.2015.05.116}
}

@article{yipMeasuringMagneticField2019,
  title = {Measuring Magnetic Field Texture in Correlated Electron Systems under Extreme Conditions},
  author = {Yip, King Yau and Ho, Kin On and Yu, King Yiu and Chen, Yang and Zhang, Wei and Kasahara, S. and Mizukami, Y. and Shibauchi, T. and Matsuda, Y. and Goh, Swee K. and Yang, Sen},
  year = 2019,
  month = dec,
  journal = {Science},
  volume = {366},
  number = {6471},
  pages = {1355--1359},
  publisher = {American Association for the Advancement of Science},
  doi = {10.1126/science.aaw4278}
}

@article{adachiPressureEffectBi22122019,
  title = {Pressure Effect in {Bi-2212} and {Bi-2223} Cuprate Superconductor},
  author = {Adachi, Shintaro and Matsumoto, Ryo and Hara, Hiroshi and Saito, Yoshito and Song, Peng and Takeya, Hiroyuki and Watanabe, Takao and Takano, Yoshihiko},
  year = {2019},
  month = mar,
  journal = {Appl. Phys. Express},
  volume = {12},
  number = {4},
  pages = {043002},
  publisher = {IOP Publishing},
  issn = {1882-0786},
  doi = {10.7567/1882-0786/ab0521},
  langid = {english}
}

@article{adachiUnscalingSuperconductingParameters2015,
  title = {Unscaling Superconducting Parameters with ${T}_{\mathrm{c}}$ for {Bi-2212} and {Bi-2223}: A Magnetotransport Study in the Superconductive Fluctuation Regime},
  author = {Adachi, Shintaro and Usui, Tomohiro and Ito, Yasuhito and Kudo, Hironobu and Kushibiki, Haruki and Murata, Kosuke and Watanabe, Takao and Kudo, Kazutaka and Nishizaki, Terukazu and Kobayashi, Norio and Kimura, Shojiro and Fujita, Masaki and Yamada, Kazuyoshi and Noji, Takashi and Koike, Yoji and Fujii, Takenori},
  year = {2015},
  month = feb,
  journal = {J. Phys. Soc. Jpn.},
  volume = {84},
  number = {2},
  pages = {024706},
  publisher = {The Physical Society of Japan},
  issn = {0031-9015},
  doi = {10.7566/JPSJ.84.024706}
}

@article{balasubramanianNanoscaleImagingMagnetometry2008a,
  title = {Nanoscale Imaging Magnetometry with Diamond Spins under Ambient Conditions},
  author = {Balasubramanian, Gopalakrishnan and Chan, I. Y. and Kolesov, Roman and {Al-Hmoud}, Mohannad and Tisler, Julia and Shin, Chang and Kim, Changdong and Wojcik, Aleksander and Hemmer, Philip R. and Krueger, Anke and Hanke, Tobias and Leitenstorfer, Alfred and Bratschitsch, Rudolf and Jelezko, Fedor and Wrachtrup, J{\"o}rg},
  year = {2008},
  month = oct,
  journal = {Nature},
  volume = {455},
  number = {7213},
  pages = {648--651},
  publisher = {Nature Publishing Group},
  issn = {1476-4687},
  doi = {10.1038/nature07278},
  langid = {english}
}

@article{bhattacharyyaImagingMeissnerEffect2024,
  title = {Imaging the {Meissner} Effect in Hydride Superconductors Using Quantum Sensors},
  author = {Bhattacharyya, P. and Chen, W. and Huang, X. and Chatterjee, S. and Huang, B. and Kobrin, B. and Lyu, Y. and Smart, T. J. and Block, M. and Wang, E. and Wang, Z. and Wu, W. and Hsieh, S. and Ma, H. and Mandyam, S. and Chen, B. and Davis, E. and Geballe, Z. M. and Zu, C. and Struzhkin, V. and Jeanloz, R. and Moore, J. E. and Cui, T. and Galli, G. and Halperin, B. I. and Laumann, C. R. and Yao, N. Y.},
  year = {2024},
  month = mar,
  journal = {Nature},
  volume = {627},
  number = {8002},
  pages = {73--79},
  publisher = {Nature Publishing Group},
  issn = {1476-4687},
  doi = {10.1038/s41586-024-07026-7},
  langid = {english}
}

@article{chenEnhancementSuperconductivityPressuredriven2010,
  title = {Enhancement of Superconductivity by Pressure-Driven Competition in Electronic Order},
  author = {Chen, Xiao-Jia and Struzhkin, Viktor V. and Yu, Yong and Goncharov, Alexander F. and Lin, Cheng-Tian and Mao, Ho-kwang and Hemley, Russell J.},
  year = {2010},
  month = aug,
  journal = {Nature},
  volume = {466},
  number = {7309},
  pages = {950--953},
  publisher = {Nature Publishing Group},
  issn = {1476-4687},
  doi = {10.1038/nature09293},
  langid = {english}
}

@article{chenHighpressurePhaseDiagram2004a,
  title = {High-Pressure Phase Diagram of {$\mathrm{Bi}_{2}\mathrm{Sr}_{2}\mathrm{CaCu}_{2}\mathrm{O}_{8+\delta}$} Single Crystals},
  author = {Chen, Xiao-Jia and Struzhkin, Viktor V. and Hemley, Russell J. and Mao, Ho-kwang and Kendziora, Chris},
  year = {2004},
  month = dec,
  journal = {Phys. Rev. B},
  volume = {70},
  number = {21},
  pages = {214502},
  publisher = {American Physical Society},
  doi = {10.1103/PhysRevB.70.214502}
}

@misc{chenImagingMagneticFlux2025a,
  title = {Imaging Magnetic Flux Trapping in Lanthanum Hydride Using Diamond Quantum Sensors},
  author = {Chen, Yang and Wen, Junyan and He, Ze-Xu and Fan, Jing-Wei and Pan, Xin-Yu and Ji, Cheng and Gou, Huiyang and Yu, Xiaohui and Chen, Liucheng and Liu, Gang-Qin},
  year = {2025},
  month = oct,
  number = {arXiv:2510.21877},
  eprint = {2510.21877},
  primaryclass = {cond-mat},
  publisher = {arXiv},
  doi = {10.48550/arXiv.2510.21877},
  archiveprefix = {arXiv}
}

@article{degenScanningMagneticField2008,
  title = {Scanning Magnetic Field Microscope with a Diamond Single-Spin Sensor},
  author = {Degen, C. L.},
  year = {2008},
  month = jun,
  journal = {Applied Physics Letters},
  volume = {92},
  number = {24},
  pages = {243111},
  issn = {0003-6951},
  doi = {10.1063/1.2943282}
}

@article{dengHigherSuperconductingTransition2019a,
  title = {Higher Superconducting Transition Temperature by Breaking the Universal Pressure Relation},
  author = {Deng, Liangzi and Zheng, Yongping and Wu, Zheng and Huyan, Shuyuan and Wu, Hung-Cheng and Nie, Yifan and Cho, Kyeongjae and Chu, Ching-Wu},
  year = {2019},
  month = feb,
  journal = {Proceedings of the National Academy of Sciences},
  volume = {116},
  number = {6},
  pages = {2004--2008},
  publisher = {Proceedings of the National Academy of Sciences},
  doi = {10.1073/pnas.1819512116}
}

@article{dohertyElectronicPropertiesMetrology2014,
  title = {Electronic Properties and Metrology Applications of the Diamond {$\mathrm{NV}^{-}$} Center under Pressure},
  author = {Doherty, Marcus W. and Struzhkin, Viktor V. and Simpson, David A. and McGuinness, Liam P. and Meng, Yufei and Stacey, Alastair and Karle, Timothy J. and Hemley, Russell J. and Manson, Neil B. and Hollenberg, Lloyd C. L. and Prawer, Steven},
  year = {2014},
  month = jan,
  journal = {Phys. Rev. Lett.},
  volume = {112},
  number = {4},
  pages = {047601},
  publisher = {American Physical Society},
  doi = {10.1103/PhysRevLett.112.047601}
}

@article{fujiiDopingDependenceAnisotropic2002,
  title = {Doping Dependence of Anisotropic Resistivities in the Trilayered Superconductor {$\mathrm{Bi}_{2}\mathrm{Sr}_{2}\mathrm{Ca}_{2}\mathrm{Cu}_{3}\mathrm{O}_{10+\delta}$}},
  author = {Fujii, Takenori and Terasaki, Ichiro and Watanabe, Takao and Matsuda, Azusa},
  year = {2002},
  month = jul,
  journal = {Phys. Rev. B},
  volume = {66},
  number = {2},
  pages = {024507},
  publisher = {American Physical Society},
  doi = {10.1103/PhysRevB.66.024507}
}

@article{fujiiSinglecrystalGrowthBi2Sr2Ca2Cu3O10+d2001,
  title = {Single-Crystal Growth of {$\mathrm{Bi}_{2}\mathrm{Sr}_{2}\mathrm{Ca}_{2}\mathrm{Cu}_{3}\mathrm{O}_{10+\delta}$} ({Bi-2223}) by {TSFZ} Method},
  author = {Fujii, Takenori and Watanabe, Takao and Matsuda, Azusa},
  year = {2001},
  month = feb,
  journal = {Journal of Crystal Growth},
  volume = {223},
  number = {1},
  pages = {175--180},
  issn = {0022-0248},
  doi = {10.1016/S0022-0248(00)01028-9}
}

@article{gruberScanningConfocalOptical1997a,
  title = {Scanning Confocal Optical Microscopy and Magnetic Resonance on Single Defect Centers},
  author = {Gruber, A. and Dr{\"a}benstedt, A. and Tietz, C. and Fleury, L. and Wrachtrup, J. and von Borczyskowski, C.},
  year = {1997},
  month = jun,
  journal = {Science},
  volume = {276},
  number = {5321},
  pages = {2012--2014},
  publisher = {American Association for the Advancement of Science},
  doi = {10.1126/science.276.5321.2012}
}

@misc{haoDiamondQuantumSensing2025,
  title = {Diamond Quantum Sensing at Record High Pressure up to 240 {GPa}},
  author = {Hao, Qingtao and He, Ze-Xu and Zuo, Na and Chen, Yang and Xing, Xiangzhuo and Zhang, Xiaoran and Zhuang, Xinyu and Shi, Zhixiang and Chen, Xin and Guo, Jian-Gang and Liu, Gang-Qin and Liu, Xiaobing and Ma, Yanming},
  year = {2025},
  month = oct,
  number = {arXiv:2510.26605},
  eprint = {2510.26605},
  primaryclass = {quant-ph},
  publisher = {arXiv},
  doi = {10.48550/arXiv.2510.26605},
  archiveprefix = {arXiv}
}

@article{hsiehImagingStressMagnetism2019,
  title = {Imaging Stress and Magnetism at High Pressures Using a Nanoscale Quantum Sensor},
  author = {Hsieh, S. and Bhattacharyya, P. and Zu, C. and Mittiga, T. and Smart, T. J. and Machado, F. and Kobrin, B. and H{\"o}hn, T. O. and Rui, N. Z. and Kamrani, M. and Chatterjee, S. and Choi, S. and Zaletel, M. and Struzhkin, V. V. and Moore, J. E. and Levitas, V. I. and Jeanloz, R. and Yao, N. Y.},
  year = {2019},
  month = dec,
  journal = {Science},
  volume = {366},
  number = {6471},
  pages = {1349--1354},
  publisher = {American Association for the Advancement of Science},
  doi = {10.1126/science.aaw4352}
}

@article{lesikMagneticMeasurementsMicrometersized2019,
  title = {Magnetic Measurements on Micrometer-Sized Samples under High Pressure Using Designed {NV} Centers},
  author = {Lesik, Margarita and Plisson, Thomas and Toraille, Lo{\"i}c and Renaud, Justine and Occelli, Florent and Schmidt, Martin and Salord, Olivier and Delobbe, Anne and Debuisschert, Thierry and Rondin, Lo{\"i}c and Loubeyre, Paul and Roch, Jean-Fran{\c c}ois},
  year = {2019},
  month = dec,
  journal = {Science},
  volume = {366},
  number = {6471},
  pages = {1359--1362},
  publisher = {American Association for the Advancement of Science},
  doi = {10.1126/science.aaw4329}
}

@article{liuCoherentQuantumControl2019,
  title = {Coherent Quantum Control of Nitrogen-Vacancy Center Spins near 1000 {Kelvin}},
  author = {Liu, Gang-Qin and Feng, Xi and Wang, Ning and Li, Quan and Liu, Ren-Bao},
  year = {2019},
  month = mar,
  journal = {Nat. Commun.},
  volume = {10},
  number = {1},
  pages = {1344},
  publisher = {Nature Publishing Group},
  issn = {2041-1723},
  doi = {10.1038/s41467-019-09327-2},
  langid = {english}
}

@article{maoSolidsLiquidsGases2018,
  title = {Solids, Liquids, and Gases under High Pressure},
  author = {Mao, Ho-Kwang and Chen, Xiao-Jia and Ding, Yang and Li, Bing and Wang, Lin},
  year = {2018},
  month = mar,
  journal = {Rev. Mod. Phys.},
  volume = {90},
  number = {1},
  pages = {015007},
  publisher = {American Physical Society},
  doi = {10.1103/RevModPhys.90.015007}
}

@article{markProgressProspectsCuprate2022,
  title = {Progress and Prospects for Cuprate High Temperature Superconductors under Pressure},
  author = {Mark, Alexander C. and Campuzano, Juan Carlos and Hemley, Russell J.},
  year = {2022},
  month = apr,
  journal = {High Pressure Research},
  volume = {42},
  number = {2},
  pages = {137--199},
  publisher = {Taylor \& Francis},
  issn = {0895-7959},
  doi = {10.1080/08957959.2022.2059366},
  langid = {american}
}

@article{markStructureEquationState2023,
  title = {Structure and Equation of State of {$\mathrm{Bi}_{2}\mathrm{Sr}_{2}\mathrm{Ca}_{n-1}\mathrm{Cu}_{n}\mathrm{O}_{2n+4+\delta}$} from X-Ray Diffraction to Megabar Pressures},
  author = {Mark, Alexander C. and Ahart, Muhtar and Kumar, Ravhi and Park, Changyong and Meng, Yue and Popov, Dmitry and Deng, Liangzi and Chu, Ching-Wu and Campuzano, Juan Carlos and Hemley, Russell J.},
  year = {2023},
  month = jun,
  journal = {Phys. Rev. Mater.},
  volume = {7},
  number = {6},
  pages = {064803},
  publisher = {American Physical Society},
  doi = {10.1103/PhysRevMaterials.7.064803}
}

@article{matsumotoCrystalGrowthHighPressure2021,
  title = {Crystal Growth and High-Pressure Effects of {Bi-Based} Superconducting Whiskers},
  author = {Matsumoto, Ryo and Yamamoto, Sayaka and Takano, Yoshihiko and Tanaka, Hiromi},
  year = {2021},
  month = may,
  journal = {ACS Omega},
  volume = {6},
  number = {18},
  pages = {12179--12186},
  publisher = {American Chemical Society},
  doi = {10.1021/acsomega.1c00880}
}

@article{mazeNanoscaleMagneticSensing2008,
  title = {Nanoscale Magnetic Sensing with an Individual Electronic Spin in Diamond},
  author = {Maze, J. R. and Stanwix, P. L. and Hodges, J. S. and Hong, S. and Taylor, J. M. and Cappellaro, P. and Jiang, L. and Dutt, M. V. Gurudev and Togan, E. and Zibrov, A. S. and Yacoby, A. and Walsworth, R. L. and Lukin, M. D.},
  year = {2008},
  month = oct,
  journal = {Nature},
  volume = {455},
  number = {7213},
  pages = {644--647},
  publisher = {Nature Publishing Group},
  issn = {1476-4687},
  doi = {10.1038/nature07279},
  langid = {english}
}

@article{mitoUniaxialStrainEffects2016,
  title = {Uniaxial Strain Effects on Superconducting Transition in {$\mathrm{Y}_{0.98}\mathrm{Ca}_{0.02}\mathrm{Ba}_{2}\mathrm{Cu}_{4}\mathrm{O}_{8}$}},
  author = {Mito, Masaki and Goto, Hiroki and Matsui, Hideaki and Deguchi, Hiroyuki and Matsumoto, Kaname and Hara, Hiroshi and Ozaki, Toshinori and Takeya, Hiroyuki and Takano, Yoshihiko},
  year = {2016},
  month = feb,
  journal = {J. Phys. Soc. Jpn.},
  volume = {85},
  number = {2},
  pages = {024711},
  publisher = {The Physical Society of Japan},
  issn = {0031-9015},
  doi = {10.7566/JPSJ.85.024711}
}

@article{mitoUniaxialStrainEffects2017,
  title = {Uniaxial Strain Effects on the Superconducting Transition in {Re-doped} {Hg-1223} Cuprate Superconductors},
  author = {Mito, Masaki and Ogata, Kazuma and Goto, Hiroki and Tsuruta, Kazuki and Nakamura, Kazuma and Deguchi, Hiroyuki and Horide, Tomoya and Matsumoto, Kaname and Tajiri, Takayuki and Hara, Hiroshi and Ozaki, Toshinori and Takeya, Hiroyuki and Takano, Yoshihiko},
  year = {2017},
  month = feb,
  journal = {Phys. Rev. B},
  volume = {95},
  number = {6},
  pages = {064503},
  publisher = {American Physical Society},
  doi = {10.1103/PhysRevB.95.064503}
}

@article{nokelainenSecondDomeSuperconductivity2024,
  title = {Second Dome of Superconductivity in {$\mathrm{YBa}_{2}\mathrm{Cu}_{3}\mathrm{O}_{7}$} at High Pressure},
  author = {Nokelainen, Johannes and Matzelle, Matthew E. and Lane, Christopher and Atlam, Nabil and Zhang, Ruiqi and Markiewicz, Robert S. and Barbiellini, Bernardo and Sun, Jianwei and Bansil, Arun},
  year = {2024},
  month = jul,
  journal = {Phys. Rev. B},
  volume = {110},
  number = {2},
  pages = {L020502},
  publisher = {American Physical Society},
  doi = {10.1103/PhysRevB.110.L020502}
}

@article{oortOpticallyDetectedSpin1988,
  title = {Optically Detected Spin Coherence of the Diamond {N-V} Centre in Its Triplet Ground State},
  author = {van Oort, E. and Manson, N. B. and Glasbeek, M.},
  year = {1988},
  month = aug,
  journal = {J. Phys. C: Solid State Phys.},
  volume = {21},
  number = {23},
  pages = {4385},
  issn = {0022-3719},
  doi = {10.1088/0022-3719/21/23/020},
  langid = {english}
}

@article{sakakibaraTheoreticalAnalysisPossibility2024,
  title = {Theoretical Analysis on the Possibility of Superconductivity in the Trilayer {Ruddlesden-Popper} Nickelate {$\mathrm{La}_{4}\mathrm{Ni}_{3}\mathrm{O}_{10}$} under Pressure and Its Experimental Examination: Comparison with {$\mathrm{La}_{3}\mathrm{Ni}_{2}\mathrm{O}_{7}$}},
  author = {Sakakibara, Hirofumi and Ochi, Masayuki and Nagata, Hibiki and Ueki, Yuta and Sakurai, Hiroya and Matsumoto, Ryo and Terashima, Kensei and Hirose, Keisuke and Ohta, Hiroto and Kato, Masaki and Takano, Yoshihiko and Kuroki, Kazuhiko},
  year = {2024},
  month = apr,
  journal = {Phys. Rev. B},
  volume = {109},
  number = {14},
  pages = {144511},
  publisher = {American Physical Society},
  doi = {10.1103/PhysRevB.109.144511}
}

@article{schaefer-nolteDiamondbasedScanningProbe2014,
  title = {A Diamond-Based Scanning Probe Spin Sensor Operating at Low Temperature in Ultra-High Vacuum},
  author = {{Schaefer-Nolte}, E. and Reinhard, F. and Ternes, M. and Wrachtrup, J. and Kern, K.},
  year = {2014},
  month = jan,
  journal = {Rev. Sci. Instrum.},
  volume = {85},
  number = {1},
  pages = {013701},
  issn = {0034-6748},
  doi = {10.1063/1.4858835}
}

@article{scheideggerScanningNitrogenvacancyMagnetometry2022a,
  title = {Scanning Nitrogen-Vacancy Magnetometry down to 350 {mK}},
  author = {Scheidegger, P. J. and Diesch, S. and Palm, M. L. and Degen, C. L.},
  year = {2022},
  month = may,
  journal = {Applied Physics Letters},
  volume = {120},
  number = {22},
  pages = {224001},
  issn = {0003-6951},
  doi = {10.1063/5.0093548}
}

@article{shimizuPressureinducedSuperconductivityElemental2005,
  title = {Pressure-Induced Superconductivity in Elemental Materials},
  author = {Shimizu, Katsuya and Amaya, Kiichi and Suzuki, Naoshi},
  year = {2005},
  month = may,
  journal = {J. Phys. Soc. Jpn.},
  volume = {74},
  number = {5},
  pages = {1345--1357},
  publisher = {The Physical Society of Japan},
  issn = {0031-9015},
  doi = {10.1143/JPSJ.74.1345}
}

@article{sunSignaturesSuperconductivity802023a,
  title = {Signatures of Superconductivity near 80 {K} in a Nickelate under High Pressure},
  author = {Sun, Hualei and Huo, Mengwu and Hu, Xunwu and Li, Jingyuan and Liu, Zengjia and Han, Yifeng and Tang, Lingyun and Mao, Zhongquan and Yang, Pengtao and Wang, Bosen and Cheng, Jinguang and Yao, Dao-Xin and Zhang, Guang-Ming and Wang, Meng},
  year = {2023},
  month = sep,
  journal = {Nature},
  volume = {621},
  number = {7979},
  pages = {493--498},
  publisher = {Nature Publishing Group},
  issn = {1476-4687},
  doi = {10.1038/s41586-023-06408-7},
  langid = {english}
}

@article{takeshitaZeroResistivity1502013,
  title = {Zero Resistivity above 150 {K} in {$\mathrm{HgBa}_{2}\mathrm{Ca}_{2}\mathrm{Cu}_{3}\mathrm{O}_{8+\delta}$} at High Pressure},
  author = {Takeshita, Nao and Yamamoto, Ayako and Iyo, Akira and Eisaki, Hiroshi},
  year = {2013},
  month = feb,
  journal = {J. Phys. Soc. Jpn.},
  volume = {82},
  number = {2},
  pages = {023711},
  publisher = {The Physical Society of Japan},
  issn = {0031-9015},
  doi = {10.7566/JPSJ.82.023711}
}

@article{taylorHighsensitivityDiamondMagnetometer2008a,
  title = {High-Sensitivity Diamond Magnetometer with Nanoscale Resolution},
  author = {Taylor, J. M. and Cappellaro, P. and Childress, L. and Jiang, L. and Budker, D. and Hemmer, P. R. and Yacoby, A. and Walsworth, R. and Lukin, M. D.},
  year = {2008},
  month = oct,
  journal = {Nature Phys.},
  volume = {4},
  number = {10},
  pages = {810--816},
  publisher = {Nature Publishing Group},
  issn = {1745-2481},
  doi = {10.1038/nphys1075},
  langid = {english}
}

@article{toyliMeasurementControlSingle2012a,
  title = {Measurement and Control of Single Nitrogen-Vacancy Center Spins above 600 {K}},
  author = {Toyli, D. M. and Christle, D. J. and Alkauskas, A. and Buckley, B. B. and {Van de Walle}, C. G. and Awschalom, D. D.},
  year = {2012},
  month = jul,
  journal = {Phys. Rev. X},
  volume = {2},
  number = {3},
  pages = {031001},
  publisher = {American Physical Society},
  doi = {10.1103/PhysRevX.2.031001}
}

@article{wangImagingMagneticTransition2024,
  title = {Imaging Magnetic Transition of Magnetite to Megabar Pressures Using Quantum Sensors in Diamond Anvil Cell},
  author = {Wang, Mengqi and Wang, Yu and Liu, Zhixian and Xu, Ganyu and Yang, Bo and Yu, Pei and Sun, Haoyu and Ye, Xiangyu and Zhou, Jingwei and Goncharov, Alexander F. and Wang, Ya and Du, Jiangfeng},
  year = {2024},
  month = oct,
  journal = {Nat. Commun.},
  volume = {15},
  number = {1},
  pages = {8843},
  publisher = {Nature Publishing Group},
  issn = {2041-1723},
  doi = {10.1038/s41467-024-52272-y},
  langid = {english}
}

@article{wangPressureInducedSuperconductivityPolycrystalline2024,
  title = {Pressure-Induced Superconductivity In Polycrystalline {$\mathrm{La}_{3}\mathrm{Ni}_{2}\mathrm{O}_{7-\delta}$}},
  author = {Wang, G. and Wang, N. N. and Shen, X. L. and Hou, J. and Ma, L. and Shi, L. F. and Ren, Z. A. and Gu, Y. D. and Ma, H. M. and Yang, P. T. and Liu, Z. Y. and Guo, H. Z. and Sun, J. P. and Zhang, G. M. and Calder, S. and Yan, J.-Q. and Wang, B. S. and Uwatoko, Y. and Cheng, J.-G.},
  year = {2024},
  month = mar,
  journal = {Phys. Rev. X},
  volume = {14},
  number = {1},
  pages = {011040},
  publisher = {American Physical Society},
  doi = {10.1103/PhysRevX.14.011040}
}

@article{yamamotoHighPressureEffects2015,
  title = {High Pressure Effects Revisited for the Cuprate Superconductor Family with Highest Critical Temperature},
  author = {Yamamoto, Ayako and Takeshita, Nao and Terakura, Chieko and Tokura, Yoshinori},
  year = {2015},
  month = dec,
  journal = {Nat. Commun.},
  volume = {6},
  number = {1},
  pages = {8990},
  publisher = {Nature Publishing Group},
  issn = {2041-1723},
  doi = {10.1038/ncomms9990},
  langid = {english}
}

@article{yamanakaSiliconClathratesCarbon2010,
  title = {Silicon Clathrates and Carbon Analogs: High Pressure Synthesis, Structure, and Superconductivity},
  author = {Yamanaka, Shoji},
  year = {2010},
  month = feb,
  journal = {Dalton Trans.},
  volume = {39},
  number = {8},
  pages = {1901--1915},
  publisher = {The Royal Society of Chemistry},
  issn = {1477-9234},
  doi = {10.1039/B918480E},
  langid = {english}
}

\end{document}